\documentclass[12pt]{article}
\usepackage{array}
\usepackage{graphicx}
\usepackage{amssymb}
\usepackage{amsmath}
\usepackage{multirow}
\input{epsf}
\usepackage{cite}
\def\@fmsl@sh#1#2#3{\m@th\ooalign{$\hfil#1\mkern#2/\hfil$\crcr$#1#3$}}
 \def\eq#1\en{\begin{equation}#1\end{equation}}
\def\s[#1,#2]{[#1\stackrel{\star}{,}#2]}
\def\sx[#1,#2]{[#1\stackrel{\star_{x}}{,}#2]}

\newcommand{\nc}{\newcommand}
\nc{\beq}{\begin{equation}}
\nc{\eeq}{\end{equation}}
\nc{\beqa}{\begin{eqnarray}}
\nc{\eeqa}{\end{eqnarray}}

\def\bc{\begin{center}}
\def\ec{\end{center}}

\def\to{\rightarrow}

\def\gsim{\mathrel{\mathpalette\atversim>}}

\def\bc{\begin{center}}
\def\ec{\end{center}}

\def\gsim{\mathrel{\rlap{\lower4pt\hbox{\hskip1pt$\sim$}}

    \raise1pt\hbox{$>$}}}       

\def\gsim{\mathrel{\rlap{\lower4pt\hbox{\hskip1pt$\sim$}}
    \raise1pt\hbox{$>$}}}       

\begin{document}
\makeatletter
\def\fmslash{\@ifnextchar[{\fmsl@sh}{\fmsl@sh[0mu]}}
\def\fmsl@sh[#1]#2{%
  \mathchoice
    {\@fmsl@sh\displaystyle{#1}{#2}}%
    {\@fmsl@sh\textstyle{#1}{#2}}%
    {\@fmsl@sh\scriptstyle{#1}{#2}}%
    {\@fmsl@sh\scriptscriptstyle{#1}{#2}}}
\def\@fmsl@sh#1#2#3{\m@th\ooalign{$\hfil#1\mkern#2/\hfil$\crcr$#1#3$}}
\makeatother

\thispagestyle{empty}
\begin{titlepage}
\boldmath
\begin{center}
  \Large {\bf Non thermal small black holes}
    \end{center}
\unboldmath
\vspace{0.2cm}
\begin{center}
{ {\large Xavier Calmet}\footnote{x.calmet@sussex.ac.uk},
{\large Dionysios Fragkakis} \footnote{d.fragkakis@sussex.ac.uk}
 and
  {\large Nina Gausmann} \footnote{n.gausmann@sussex.ac.uk}
}
 \end{center}
\begin{center}
{\sl Physics and Astronomy, 
University of Sussex,   Falmer, Brighton, BN1 9QH, UK 
}
\end{center}
\vspace{\fill}
\begin{abstract}
\noindent
In this chapter we review the current theoretical state of the art of small black holes at the LHC. We discuss the production mechanism for small non thermal black holes at the LHC and discuss new signatures due to a possible discrete mass spectrum of these black holes.

\end{abstract}  
\end{titlepage}



\newpage

One of the major remaining challenges of theoretical physics is to understand gravity at the quantum level. While quantum gravity is most likely impossible to probe experimentally  in conventional models of short distance space-time where quantum effects are expected to become relevant at energies of some $10^{19}$ GeV, there is a class of models with low scale quantum gravity which is being probed directly by measurements at the Large Hadron Collider (LHC) at CERN which predict that quantum gravitational effects could become important at a few TeVs. These models require the existence of a large extra-dimensional volume \cite{ArkaniHamed:1998rs,Randall:1999ee} or a large number of particles which lead to a running of the Planck mass \cite{Calmet:2008tn}, see e.g. \cite{Calmet:2010nt} for a recent review. The production of small black holes at the LHC would be one of the most intriguing signals of low scale quantum gravitational physics. We emphasize that recent studies have shown that brane world  models as well as models with a large number of particles typically have unitarity problems well below the effective Planck mass and new physics effects, responsible for fixing the perturbative unitarity of the S-matrix, should first be observed before quantum gravitational effects are seen at the LHC \cite{Atkins:2010eq,Atkins:2010re,Antoniadis:2011bi}.

Black holes are fascinating objects because they involve physics under extreme conditions. Indeed, if the scale of quantum gravity is truly as low as a few TeV, the most striking feature of these models is the prediction that colliders such as the LHC may be able to create small black holes \cite{Dimopoulos:2001hw,Banks:1999gd,Giddings:2001bu,Feng:2001ib,Anchordoqui:2003ug,Anchordoqui:2001cg,Anchordoqui:2003jr,Meade:2007sz,Calmet:2008dg,Calmet:2011ta,Calmet:2010vp} which would allow us to probe quantum gravity directly in an experiment.

It is important to realize that if produced at the LHC, these black holes would be quite different from astrophysical black holes in the sense that their masses would be close to the Planck mass, i.e., in these scenarios, a few TeVs. Depending on the ratio of their masses to the Planck mass, we shall refer to these small black holes as semi-classical black holes or quantum black holes.

The last ten years have resulted in impressive progress in our understanding of the production of small black holes in the collision of two particles at very high energy.
In the early days of black hole formation at colliders, the hoop conjecture \cite{hoop} due to Kip Thorne was used as a criteria for gravitational collapse in the collision of two particles head to head. The hoop conjecture states that if an amount of energy $E$ is confined to a spherical region of space-time with a radius $R$ with $R < E$, then that region will eventually evolve into a black hole.  Natural units were used: $\hbar, c$ and Newton's constant are set to unity. While the hoop conjecture is to a certain extend at the  hand waving level, there is  a convincing proof of gravitational collapse in the case of a collision of two particles with non zero impact parameter.  It had been known since the works of Penrose (unpublished) and later on by D'Eath and Payne \cite{D'Eath:1992hb}  that a small black hole will be formed in the head to head collision of two particles with zero impact parameter. However, the relevant case for the LHC is that of a non vanishing impact parameter. The resolution of the problem was given by Eardley and Giddings \cite{Eardley:2002re}, see also \cite{Hsu:2002bd}, who were able to construct a closed trapped surface. Apparently Penrose had also derived the latter result, but it was never published.

The construction of Eardley and Giddings  \cite{Eardley:2002re} is valid in the limit where the mass of the black holes and hence the center of mass energy is much larger than the effective reduced Planck mass. In other words, the construction applies to two colliding planets with a very high center of mass energy tending towards infinity. The black hole formed in that limit is essentially classical. It was shown by Hsu \cite{Hsu:2002bd}, using a path integral formulation, that this construction could be extended to the semi-classical regime if the small black hole mass is somewhat larger than the Planck mass The ratio between the first semi-classical black hole mass and that of the Planck mass can be estimated. In the case of ADD, it is typically taken to be of the order of 5, while it could easily be 20 for RS \cite{Meade:2007sz}.  Semi-classical black holes are thermal objects that are expected to decay via Hawking radiation to many particles, typically of the order of 20, after a spin down phase. This final explosion would lead to  a spectacular signature in a detector. It is however now well understood \cite{Anchordoqui:2003ug,Meade:2007sz,Calmet:2008dg}  that it is very unlikely that semi-classical black holes will be produced at the LHC because the center of mass energy is not high enough. The main reasons are that not all the energy of the partons is available for black hole formation and the parton distribution functions tend to fall off very fast.

It has been proposed in  \cite{Calmet:2008dg} to extrapolate the semi-classical black hole into the quantum regime of quantum gravity and to consider  the production of quantum black holes (QBHs)  at the LHC. QBHs are defined as the quantum analogs of ordinary black holes as their mass and Schwarzschild radius approach the quantum gravity scale. QBHs do not have semi-classical space-times and are not necessarily well-described by the usual Hawking temperature or black hole thermodynamics. In other words, they are non thermal. In many respects they are perhaps more analogous to strongly coupled resonances or bound states than to large black holes. QBHs presumably decay only to a few particles, each with Compton wavelength of order the size of the QBH. It seems unlikely that they would decay to a much larger number of longer wavelength modes. 

An important question is whether the mass of the quantum black holes is quantized or continuous as expected in the case of macroscopic black holes. Most studies so far are assuming that QBHs have a continuous mass spectrum despite some recent warnings that the quantum black hole masses ought to be quantized \cite{Dvali:2011nh}. We shall now describe the production cross section of quantum black holes at the LHC. We shall first discuss the continuous mass spectrum and then the discrete mass spectrum. In both cases, we shall assume that the cross sections can be extrapolated from the cross section obtained for semiclassical black holes, i.e. the geometrical cross section $\pi r_s^2$ where $r_s$ is the Schwarzschild radius.

{\it Continuous mass spectrum}\\
In that case, the LHC production cross section for QBHs with a continuous mass spectrum is assumed to be of the form
\begin{eqnarray}
\sigma^{pp}(s,x_{min},n,M_D) &=& \int_0^1 2z dz \int_{\frac{(x_{min} M_D)^2}{y(z)^2 s}}^1 du \int_u^1 \frac{dv}{v}  \\ \nonumber && \times F(n) \pi r_s^2(us,n,M_D) \sum_{i,j} f_i(v,Q) f_j(u/v,Q)
\end{eqnarray}
where $M_D$ is the n dimensional reduced Planck mass, $z=b/b_{max}$, $x_{min}=M_{BH,min}/M_D$,  $n$ is the number of extra-dimensions, $F(n)$ and $y(z)$ are the factors introduced by Eardley and Giddings and by Yoshino and Nambu \cite{Yoshino:2002tx}. The $n$ dimensional  Schwarzschild radius is given by 
\begin{eqnarray}
r_s(us,n,M_D)=k(n)M_D^{-1}[\sqrt{us}/M_D]^{1/(1+n)}
\end{eqnarray}
where
\begin{eqnarray}
k(n) =  \left [2^n \sqrt{\pi}^{n-3} \frac{\Gamma((3+n)/2)}{2+n} \right ]^{1/(1+n)}.
\end{eqnarray}
The fact that these QBHs are non thermal is reflected in the assumption that they decay only to a few particles immediately after their creation.

{\it Discrete mass spectrum} \\
We now consider the discrete mass spectrum case. The cross section is given by
\begin{eqnarray}
\sigma^{pp}_{QBH}(s,M_{QBH},n,M_D) &=&  \pi r_s^2(M_{QBH},n,M_D) \int_0^1 2z dz \int_{\frac{(M_{QBH})^2}{y(z)^2 s}}^1 du \int_u^1 \frac{dv}{v}  \\ \nonumber && \times F(n)  \sum_{i,j} f_i(v,Q) f_j(u/v,Q)
\end{eqnarray}
with the constant Schwarzschild radius given by
\begin{eqnarray}
r_s(M^2_{QBH},n,M_D)=k(n)M_D^{-1}[\sqrt{M^2_{QBH}}/M_D]^{1/(1+n)}
\end{eqnarray}
where as previously
\begin{eqnarray}
k(n) =  \left [2^n \sqrt{\pi}^{n-3} \frac{\Gamma((3+n)/2)}{2+n} \right ]^{1/(1+n)}.
\end{eqnarray}
Note that the parton level cross section is constant in that case. The physics of QBHs with a discrete mass spectrum is very different from the continuous case. They are expected to behave as heavy resonances that will decay to a few particles. The total QBH cross section is given by the sum of the individual QBH production cross sections:
\begin{eqnarray}
\sigma^{pp}_{tot}(s,n,M_D) &=& \sum_i \sigma^{pp}_{QBH}(s,M^i_{QBH},n,M_D). 
\end{eqnarray}
We expect that the mass spectrum is quantized in terms of the Planck mass because of the existence of a minimal length \cite{min,Calmet:2004mp} in models incorporating quantum mechanics and general relativity. For a Planck mass at 1 TeV, one might expect 5 QBH states between the Planck mass and the semi-classical regime.

{\it Decay modes}

It is  assumed that QBHs are defined by three quantities: their mass, spin and gauge charges. Importantly, QBHs can have a QCD, or color, charge. This is not in contradiction with confinement since the typical length scale of QCD, i.e., a Fermi, is much larger than the size of a QBH, e.g., TeV$^{-1}$. The formation and decay of a QBH takes place over a small space-time region -- from the QCD perspective it is a short distance process, and hadronization occurs only subsequently.  Their decomposition modes depend on a few assumptions which are as follows
\begin{itemize}
\item[I)] Processes involving QBHs conserve QCD and U(1) charges since local gauge symmetries are not violated by gravity. Note that no similar assumption is made about global charges.
\item[II)] QBH coupling to long wavelength and highly off-shell perturbative modes is suppressed.
\end{itemize}
Assumption (II) is necessary so that precision measurements, or, possibly, proton decay do not force the quantum gravity scale to be much larger than the TeV range. It is not implausible that a nonperturbative QBH state couples only weakly to long distance or highly off-shell modes, but strongly to modes of size and energy similar to that of the hole. This is analogous to results obtained for (B+L) violating processes in the standard model: (B+L) violation is exponentially small in low energy reactions, but of order one for energies above the sphaleron mass.

Generically speaking, QBHs form representations of SU(3)$_c$ and carry a QED charge. The process of two partons $p_i$, $p_j$ forming a quantum black hole in the $c$ representation of SU(3)$_c$ and charge $q$ as: $p_i+p_j \to$ QBH$_c^q$ is considered in  \cite{Calmet:2008dg}. The following different transitions are possible at a proton collider:
\begin{itemize}
\item[a)] ${\bf 3} \times {\bf \overline 3}= {\bf 8} + {\bf 1}$ \\
\item[b)] ${\bf 3} \times {\bf 3}= {\bf 6} + {\bf \overline 3}$\\
\item[c)] ${\bf 3} \times {\bf 8}= {\bf 3} + {\bf \overline 6}+ {\bf 15}$\\
\item[d)] ${\bf 8} \times {\bf 8}= {\bf 1}_S + {\bf 8}_S+ {\bf 8}_A+{\bf 10} + {\bf \overline{10}}_A+ {\bf 27}_S$
\end{itemize}
Most of the time the black holes which are created carry a SU(3)$_c$ charge and come in different representations of SU(3)$_c$ as well as QED charges. This allows to predict how they will be produced or decay.  For example the production cross-section of a QBH$_1^0$ is given by
\begin{eqnarray}
\sigma^{pp}(s,x_{min},n,M_D) &=& \int_0^1 2z dz \int_{\frac{(x_{min} M_D)^2}{y(z)^2 s}}^1 du \int_u^1 \frac{dv}{v}  \\ \nonumber && \times F(n) \pi r_s^2(us,n,M_D)
\\ \nonumber &&
\left( \frac{1}{9} \sum_{i,j=q,{\bar q}} f_i(v,Q) f_{\bar j}(u/v,Q)
+\frac{1}{64}  f_g(v,Q) f_g(u/v,Q) \right)
\end{eqnarray}
where $i,j$ runs over all the quarks and anti-quarks subject to the constraint of QED charge neutrality,
and $f_q, f_g$ are the quark and gluon parton distribution functions. For the production of a specific member (i.e., with specified color) of the octet QBH$_8^0$, one finds the same expression. In case of a discrete mass spectrum $us$ in $r_s$ should be replaced by $M_{QBH}^2$. The details of the final states have been considered elsewhere \cite{Calmet:2008dg} and applies to the discrete mass spectrum case as well.

In conclusion, we have reviewed the production mechanism for small non thermal black holes at the LHC and discussed new signatures due to a possible discrete mass spectrum of these black holes.

\bigskip

{\it Acknowledgments:}
This work is supported in part by the European Cooperation in Science and Technology (COST) action MP0905 ``Black Holes in a Violent  Universe".  The work of NG is supported by a SEPnet PhD fellowship. 

\bigskip
\bigskip


\end{document}